 \documentclass[final,5p,times,twocolumn,number]{elsarticle}

\usepackage{amssymb,amsmath}

\usepackage{multirow}
\usepackage{hyperref}
\usepackage{subfigure}
\usepackage{mathrsfs}
\usepackage{acronym}
\usepackage[usenames, dvipsnames, svgnames]{xcolor}

\newcommand{\paper}{paper}

\newcommand{\msun}{\ensuremath{\mathrm{M}_\odot}}

\journal{Physics Letters B}

\begin{document}

\begin{frontmatter}



\title{Limits on the mass of compact objects in Ho\v{r}ava-Lifshitz gravity}


\author[first]{Edwin J. Son}
\ead{eddy@nims.re.kr}
\affiliation[first]{organization={National Institute for Mathematical Sciences},
            addressline={}, 
            city={Daejeon},
            postcode={34047}, 
            state={},
            country={Republic of Korea}}

\begin{abstract}
It is known that there exist theoretical limits on the mass of compact objects in general relativity. One is the Buchdahl limit for an object with an arbitrary equation-of-state, which turns out to be the limit for an object with a uniform density. Another one is the causal limit that is stronger than the Buchdahl limit and is related to the speed of sound inside an object. Similar theoretical limits on the mass of compact objects in deformed \ac{hl} gravity are found in this \paper. Interestingly, both the uniform density limit and the sound speed limit curves converge with the horizon curve at its minimum, where a black hole becomes extremal, i.e., $M=q$, considering the \acl{ks} vacuum, which is an asymptotically flat solution in the \ac{hl} gravity.

\end{abstract}




\end{frontmatter}



\acrodef{hl}[HL]{Ho\v{r}ava-Lifshitz}
\acrodef{ks}[KS]{Kehagias-Sfetsos}
\acrodef{tov}[TOV]{Tolman-Oppenheimer-Volkoff}

\acrodef{gr}[GR]{general relativity}
\acrodef{eos}[EOS]{equation-of-state}

\acrodef{udl}[UDL]{uniform density limit}
\acrodef{ssl}[SSL]{sound speed limit}
\acresetall


\section{Introduction}

Neutron stars are highly compact objects characterized by large masses and small radii, generating strong gravitational fields in their vicinity. Recently, neutron stars of mass $\gtrsim 2\ \msun$ have been observed, where $\msun$ represents the mass of the sun~\cite{Fonseca:2021wxt,Romani:2022jhd}. The existence of such massive neutron stars is difficult to reconcile with \ac{gr}, even when employing significantly stiff \ac{eos} models. Furthermore, it has been known that there are theoretical limits on the masses of compact objects in \ac{gr}: One is the Buchdahl limit or \ac{udl}, $\mathcal{C} \equiv G_N M / c^2 R \le \mathcal{C}_\mathrm{max} = 4/9$~\cite{Buchdahl:1959pr}, and the other is the causal limit or \ac{ssl}, $M \lesssim 3\ \msun$, where a fiducial density $\rho_u = 5\times10^{14}\ \textrm{g/cm}^3$ is assumed~\cite{Rhoades:1974fn,Kalogera:1996ci,Astashenok:2021}. Here, $G_N$ is the Newton's constant, $c$ is the speed of light, and $\msun$ is the mass of the Sun.

The Buchdahl limit has also been studied in modified theories of gravity. For example, the Buchdahl limit has been slightly extended in $f(R)$ gravity, $\mathcal{C}_\mathrm{max} = (4/9) (1 + \alpha / 6)$, where $0 \le \alpha \ll 1$ represents the deformation from \ac{gr}~\cite{PhysRevD.92.064002}. In scalar-tensor gravity, the Buchdahl limit depends on a parameter $\beta$ related to a coupling function and exceeds the original Buchdahl limit, $\mathcal{C}_\mathrm{max} > 4 / 9$, for $\beta \lesssim 0.4$, and the theory even allows $\mathcal{C}_\mathrm{max} > 1 / 2$ for $\beta \lesssim 0.2$~\cite{10.1143/PTP.100.291}. In Eddington-inpired Born Infeld gravity, $\mathcal{C}_\mathrm{max} = 2(\sqrt{\bar{a} \bar{b}} + 1)/(\sqrt{\bar{a} \bar{b}} + 2)^2$, which is less than the original Buchdahl limit for $\bar{a} \bar{b} > 1$~\cite{Feng_2019}. Thus, the Buchdahl limit is not universal through the theories of gravity. As for the causal limit, there has been a study in the context of $f(R)$ gravity, and it is shown that the upper bound remains at $3\ \msun$~\cite{Astashenok:2021}. However, it is more or less expected, since the Buchdahl limit only slightly changes in $f(R)$ gravity.

On the other hand, \ac{hl} gravity is a modified gravity theory proposed as an ultraviolet complete theory of \ac{gr} by introducing an anisotropic scaling~\cite{Horava:2008ih,Horava:2009uw,Horava:2009if}. An asymptotically flat vacuum solution, \ac{ks} black hole, exists in deformed \ac{hl} gravity~\cite{Horava:2009uw,Park:2009zra} and it approximates the Schwarzschild black hole in the infrared limit~\cite{Kehagias:2009is}. In the previous work, we have found that the compact objects in \ac{hl} may be much heavier than their \ac{gr} counterparts of similar size~\cite{Kim:2018dbs}.

Motivated by this, in this \paper, we investigate the limits on the masses of compact objects in \ac{hl} gravity. We derive an explicit form of the \ac{tov} equations~\cite{Tolman:1939jz,Oppenheimer:1939ne} in \ac{hl} gravity. By solving them with an arbitrary \ac{eos}, we obtain the masses and radii of compact objects in \ac{hl}. The limits, \ac{udl} and \ac{ssl}, are found by numerical calculations.

This \paper{} is organized as follows. In Section~\ref{sec:HL}, the \ac{tov} equations are derived from \ac{hl} action.The \ac{udl} (Buchdahl limit) in \ac{hl} gravity is obtained in Section~\ref{sec:buchdahl} by assuming positive pressure and decreasing density with respect to the radial position.The \ac{ssl} (causal limit) in \ac{hl} gravity is found in Section~\ref{sec:causal} by assuming that the sound speed inside a compact object is subluminal. Finally, discussions are addressed in Section~\ref{sec:discussion}.

\section{TOV equation in Ho\v{r}ava-Lifshitz gravity}
\label{sec:HL}

\ac{hl} gravity is a ultraviolet complete version of \ac{gr} by introducing 
an anisotropic scaling between time and space, $t \to b^{z}\, t$ and $x^i \to b\, x^i$,
and Arnowitt-Deser-Misner decomposition~\cite{Horava:2008ih,Horava:2009uw,Horava:2009if}
\begin{align}
  ds^2 &= \mathcal{G}_{\mu\nu} dx^\mu dx^\nu \notag \\
  &= - N^2 c^2 dt^2 + g_{ij} (dx^i + N^i dt) (dx^j + N^j dt),
\end{align}
where $\mathcal{G}_{\mu\nu}$ is the metric of four-dimensional spacetime and $N$, $N^i$ and $g_{ij}$ are the lapse function, the shift function and the metric of three-dimensional spatial hypersurface, respectively.

Since a power-counting renormalizable theory can be built when $z \ge D$ in D+1 dimensions~\cite{Horava:2009uw,Mukohyama:2010xz}, we consider $z=3$ in 3+1 dimensions for simplicity.
\ac{hl} gravity of $z=3$ with the \emph{softly} broken detailed balance condition is given by~\cite{Horava:2009uw,Park:2009zra}
\begin{equation}
\label{action}
\begin{aligned}[b]
I_\text{HL} &= \int dt d^3x \sqrt{g} N \bigg[ \frac{2}{\kappa^2} \left( K_{ij} K^{ij} - \lambda K^2 \right) \\
  &- \frac{\kappa^2}{2\zeta^4} \left( C_{ij} - \frac{\mu\zeta^2}{2} R_{ij} \right) \left( C^{ij} - \frac{\mu\zeta^2}{2} R^{ij} \right) \\
  &+ \frac{\kappa^2\mu^2}{8(3\lambda - 1)} \left( \frac{4\lambda - 1}{4} R^2 + ( \omega - \Lambda_W ) R + 3 \Lambda_W^2 \right) \bigg],
\end{aligned}
\end{equation}
where $K_{ij} \equiv \frac{1}{2N} \left[ \dot{g}_{ij} - \nabla_i N_j - \nabla_j N_i \right]$ is the extrinsic curvature, $R_{ij}$ is the Ricci tensor in the three-dimensional spatial hypersurface, and $C^{ij} = \varepsilon^{ik\ell} \nabla_k ( R_\ell^j - (1/4) \delta_\ell^j R )$ is the Cotton-York tensor.

By identifying the fundamental constants as follows,
\begin{equation}
\label{fund:const}
c = \frac{\kappa^2}{4} \sqrt{\frac{\mu^2}{2q^2(3\lambda-1)}}, \ G_N = \frac{\kappa^2c^2}{32\pi}, \ \Lambda = -3q^2\Lambda_W^2
\end{equation}
with $\omega$ replaced by a new parameter $q = [2(\omega-\Lambda_W)]^{-1/2}>0$ of length dimension, the Einstein-Hilbert action can be recovered in the infrared limit, i.e. by keeping up to the leading order term in $R$:
\begin{equation}
\label{act:EH}
\begin{aligned}[b]
I_\text{EH} &= \frac{c^3}{16\pi G_N} \int d^4x \sqrt{-\mathcal{G}} \left[ \mathcal{R} - 2\Lambda \right] \\
  &= \frac{c^2}{16\pi G_N} \int dt d^3x \sqrt{g} N \left[ K_{ij} K^{ij} - K^2 + c^2 \left( R - 2\Lambda \right) \right]
\end{aligned}
\end{equation}
with $\lambda=1$, where $\mathcal{R}$ is the curvature scalar of four-dimensional spacetime.

For simplicity, we consider an asymptotically flat solution and set $\Lambda_W = 0$.
Then, considering a static, spherically symmetric metric ansatz,
\begin{equation}
ds^2 = - e^{2\Phi(r)} c^2 dt^2 + \frac{dr^2}{f(r)} + r^2 \left( d\theta^2 + \sin^2\theta d\phi^2 \right),
\end{equation}
and a perfect fluid $T_{\mu\nu} = (\rho+p) u_{\mu}u_{\nu} + pg_{\mu\nu}$ with a four-vector $u_\mu = (1, 0, 0, 0)$, the equations of motion are expressed as
\begin{align}
8 \pi \rho &= \frac{1}{r^2} \left( r (1-f) + q^2 \frac{(1-f)^2}{r} \right)', \label{eom:rho} \\
8 \pi p &= \frac{1}{r^4} \big[ -(1-f) \left( r^2 - q^2 (1 - f) \right) \notag \\
  &\qquad \qquad + 2 r f \left( r^2 + 2 q^2 (1-f) \right) \Phi' \big], \label{eom:p} \\
p' &= - \left( \rho + p \right) \Phi', \label{eom:cons}
\end{align}
setting $c = G_N = 1$, for convenience. Note that the equations of motion in \ac{gr} is recovered when $q \to 0$.

Next, the \ac{ks} vacuum solution to these equations for $\rho = p = 0$ is given by~\cite{Kehagias:2009is}
\begin{align}
  f &= e^{2\Phi} = 1 + (r^2 / 2q^2) \left[ 1 - \sqrt{1 + 8 q^2 M / r^3} \right] \notag \\
  &= \frac{2 [ 1 - 2 M / r + q^2 / r^2 ]}{1 + 2 q^2 / r^2 + \sqrt{1 + 8 q^2 M / r^3}},
  \label{sol:ks}
\end{align}
where $\omega$ in the original solution was replaced by $q = (2\omega)^{-1/2}$. Then, the numerator of the solution~\eqref{sol:ks} is reminiscent of the Reissner-Nordstr\"om vacuum solution and the metric $g^{rr} = f$ vanishes at $r_\pm = M \pm \sqrt{M^2 - q^2}$. It should be noted that $q$ is not a conserved quantity but a parameter related to a coupling constant $\omega$ that represents the deformation from \ac{gr} in this static, spherically symmetric and asymptotically flat geometry. The two horizons meet $r_+ = r_- = R_c \equiv q$ at the minimum $M = M_c \equiv q$ which describes the minimal black hole, and the naked singularity appears when $M < M_c$. However, we will not address this issue further, since it is out of the scope of this \paper{}.

Let us now replace the function $f(r)$ by $m(r)$ through the relation
\begin{equation}
\label{f}
  f = \frac{1}{\mathfrak{A}} \left( 1 - \frac{2 m}{r} + \frac{q^2}{r^2} \right),
\end{equation}
where $\mathfrak{A} = 2^{-1} \big[ 1 + 2q^2/r^2 + \sqrt{1 + 8q^2 \tilde{\rho}} \big]$ with $\tilde{\rho} = m / r^3$.
Then, the equations of motion \eqref{eom:rho}--\eqref{eom:cons} are simply rewritten as
\begin{align}
m' &= 4 \pi r^2 \rho, \label{eq:m} \\
p' &= - \frac{m \rho}{r^2} \frac{\mathfrak{A} (1 + p/\rho) \left[ 1+ 4 \pi \mathfrak{B} p / \tilde{\rho} - q^2 \tilde{\rho}  \right]}{\mathfrak{B} \sqrt{1 + 8q^2 \tilde{\rho}} \left[ 1 - 2 m / r + q^2/r^2 \right]}, \label{eq:p}
\end{align}
where 
$\mathfrak{B} = 2^{-1} \big[ 1 + 2q^2 \tilde{\rho} + \sqrt{1 + 8q^2 \tilde{\rho}} \big]$.
Note that the \ac{tov} equations in \ac{gr}~\cite{Tolman:1939jz,Oppenheimer:1939ne} are recovered in the limit of $q \to 0$, in which $\mathfrak{A} \to 1$ and $\mathfrak{B} \to 1$.

\section{Uniform density limit}
\label{sec:buchdahl}

Buchdahl showed that the mass of a star (a compact object) in Schwarzschild vacuum is bounded from above and a star of uniform density may reach the upper bound~\cite{Buchdahl:1959pr}. In this section, we find a similar upper bound in \ac{ks} vacuum.

\subsection{Uniform density}

Considering a star of radius $R$ and mass $M$ with a uniform density $\rho_0$ and a boundary condition $p(R)=0$, the solutions to the \ac{tov} equations are obtained as\footnote{After this \paper{} was finalized, the author recognized that these solutions for the limited case of $\alpha < 1$ and $\sigma > 0$ had been presented in ref.~\cite{MIParkTalk2010}.}
\begin{align}
  \label{eq:sol:uniform:m}
  m &= \frac{4\pi}{3} \rho_0 r^3 = M \left( \frac{r}{R} \right)^3, \\
  \label{eq:sol:uniform:p}
  \frac{p}{\rho_0} 
    &=\left\{ \begin{aligned}
      & \frac{\chi(r) - \chi(R)}{\beta \chi(R) - \chi(r)}, && \text{ for } \rho_0 + \beta p > 0, \\
      & -\frac{\chi(r) + \chi(R)}{\beta \chi(R) + \chi(r)}, && \text{ for } \rho_0 + \beta p < 0,
    \end{aligned} \right. \\
  e^{2\Phi} 
    &=\left\{ \begin{aligned}
      & \sigma \frac{\left[ \beta \chi(R) - \chi(r) \right]^2}{(\beta - 1)^2}, && \text{ for } \rho_0 + \beta p > 0, \\
      & -\sigma \frac{\left[ \beta \chi(R) + \chi(r) \right]^2}{(\beta - 1)^2}, && \text{ for } \rho_0 + \beta p < 0,
    \end{aligned} \right.
  \label{eq:sol:uniform:lapse}
\end{align}
where $\chi(r) = \left| 1 - \alpha r^2 / q^2 \right|^{1/2}$, $\sigma = \Theta(1 - \alpha R^2 / q^2)$ with Heaviside step function $\Theta(\cdot)$, $\beta = 3 (1 + \alpha) / (1 - \alpha)$ and $\alpha = 2^{-1} [ \sqrt{1 + (32 \pi / 3) q^2 \rho_0} - 1 ] > 0$ for $\rho_0 > 0$. Note that the case of $R^2 = q^2 / \alpha$ is not under consideration because it implies negative pressure, and $\beta > 3$ when $\alpha < 1$ and $\beta < -3$ when $\alpha > 1$. At the critical value of $\alpha=1$, the density is given by $\rho_0 = 3 / 4 \pi q^2$ that is nothing but the average density of the minimal black hole, $\bar{\rho}_\mathrm{MBH} = 3 M_c / 4 \pi R_c^3 \approx 3.2 \times 10^{27} (q/1\ \textrm{cm})^{-2} \ \textrm{g/cm}^3$, where the average density of a black hole $\bar{\rho}_\mathrm{BH}$ is defined as $M_\mathrm{BH} = (4 \pi / 3) \bar{\rho}_\mathrm{BH} r_+^3$.

Note that the compact objects with uniform density are distributed along a curve $M = (4 \pi / 3) \rho_0 R^3$ for a given $\rho_0$. Thus, we have three cases for solution: i) $\alpha < 1$, equivalently $\rho_0 < \bar{\rho}_\mathrm{MBH}$ and $\beta > 3$, ii) $\alpha = 1$, $\rho_0 = \bar{\rho}_\mathrm{MBH}$ and $\beta$ diverges, and iii) $\alpha > 1$, $\rho_0 > \bar{\rho}_\mathrm{MBH}$ and $\beta < -3$. Note that in the \ac{gr} limit, $q \to 0$, the average density of the minimal black hole diverges, $\bar{\rho}_\mathrm{MBH} \to \infty$, which means that the cases ii) and iii) are beyond \ac{gr}.

\begin{figure}[tbp]
\centering
\includegraphics[width=\linewidth]{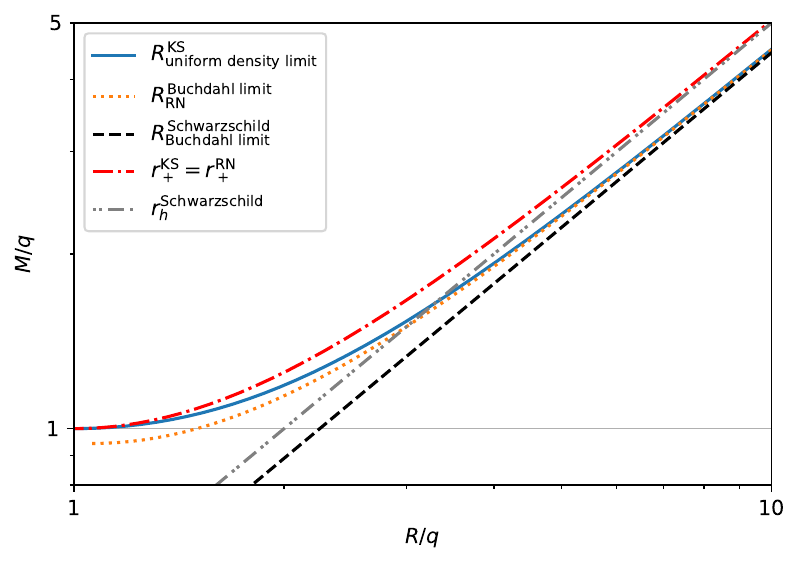}
\caption{The \acl{udl}, the upper bound of the mass $M$ of a compact object with uniform density is depicted with respect to its radius $R$, compared with the Buchdahl limit in GR for Schwarzschild and Reissner-Nordstr\"om vacua.
The horizon radii of \acl{ks}, Reissner-Nordstr\"om ($r_+^\text{\acs{ks}} = r_+^\text{RN}$) and Schwarzschild ($r_h^\text{Schwarzschild}$) black holes are also plotted.
}
\label{fig:buchdahl}
\end{figure}

First of all, for $\alpha < 1$, all curves $M = (4 \pi / 3) \rho_0 R^3$ intersect the outer horizon $r_+$ and the mass is bounded from above by $M_\text{BH} = R/2 + q^2 / 2R$. Provided that the pressure is positive, the density is bounded from above, $\rho_0 < 3 / 4 \pi q^2$, which gives the mass bound for $R < R_c$,
\begin{equation}
  \label{bound:lowmass}
  M < \frac{R^3}{q^2} .
\end{equation}

We now require a finite-pressure condition $p < \infty$ for an arbitrary $r$.
Since $\chi(r) > \chi(R)$ for $r < R$,\footnote{When $R > q / \sqrt{\alpha}$, there exists a region in $r < R$ such that $\chi(r) < \chi(R)$. However, we exclude this situation, since the pressure becomes negative.} the condition is equivalently written as $\beta \chi(R) - \chi(r) > 0$ for all $r < R$, which reduces to $\beta \chi(R) - 1 > 0$. Introducing a variable $y$ related to the mass by $y = \sqrt{1 + 8 q^2 M / R^3}$, we solve the following equation,
\begin{equation}
  y^3 + \left(1-\frac{16 q^2}{9 R^2}\right) y^2 - \left( 1 + \frac{16 q^2}{3 R^2} \right) y - 1 = 0,
\end{equation}
to find the boundary of the condition, and the solution is straightforwardly given by\footnote{The formula to solve a cubic equation can be easily found in a mathematical table, e.g., ref.~\cite{zwillinger2018crc}, or in a \texttt{Wikipedia} page~\cite{wiki:cubic_formula}.}
\begin{equation}
  y = \frac{16 - 9 R^2 / q^2  + 4 \xi \cos \left[ (\theta + 2 n \pi ) / 3 \right]}{27 R^2 / q^2},
\end{equation}
where $\theta \! = \! \cos^{-1} \! \left[ \xi^{-3} \! \left( 512 \! + \! 3024 R^2 \! / q^2 \! - \! 972 R^4 \! / q^4 \! + \! 729 R^6 \! / q^6 \right) \right]$ represents the principal value, $\xi = \sqrt{64 + 252 R^2/q^2 + 81 R^4/q^4}$, and $n$ is an arbitrary integer. Only the solution for $n = 3 k$ with an integer $k$ satisfies $\alpha = 2^{-1} ( y - 1 ) > 0$, while the other solutions, for $n = 3 k + 1$ or $n = 3 k + 2$, turn out to be negative.
Then, we have the \ac{udl} as follows:
\begin{equation}
  \label{ineq:uniform}
  M \! < \! \frac{\left( 4 \! - \! 9 R^2/q^2 \! + \! \xi \cos(\theta / 3) \right) \! \left( 8 \! + \! 9 R^2/q^2 \! + \! 2 \xi \cos(\theta / 3) \right)}{729 R/q^2} \! .
\end{equation}
The mass $M$ is bounded from above by the inequality~\eqref{ineq:uniform} and the upper bound for a given radius $R$ is depicted in Fig.~\ref{fig:buchdahl}.

In the limit of $q / R \ll 1$, the inequality~\eqref{ineq:uniform} reduces to $M / R < 4 / 9 + (16 / 27) (q / R)^2 + (64 / 729) (q / R)^4 +  O(q / R)^6$. The leading order term is nothing but the Buchdahl limit in \ac{gr} for the Schwarzschild vacuum~\cite{Buchdahl:1959pr}, which is dominant in the limit of small $q$ or large $R$. The sub-leading order term is reminiscent of the Buchdahl limit in \ac{gr} for the Reissner-Nordstr\"om vacuum~\cite{Giuliani:2007zza,Dadhich:2019jyf}, $M / R \le 4/9 + Q^2 / 2R^2$. Note that the numeric factor of the sub-leading order term is different from the Reissner-Nordstr\"om Buchdahl limit, even though the horizon is exactly same for $Q=q$.

\begin{figure*}[tb]
  \centering
  \subfigure[\ uniform density]{
    \label{fig:uniform}
    \includegraphics[width=\columnwidth]{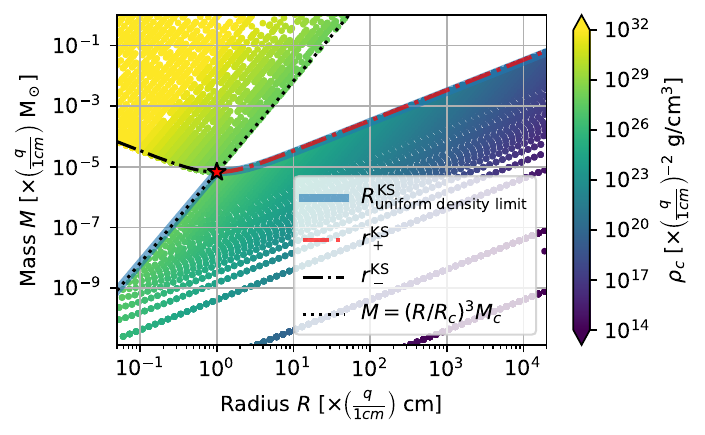}
  }
  \subfigure[\ uniform density (near the minimal black hole)]{
    \label{fig:uniform:zoom}
    \includegraphics[width=\columnwidth]{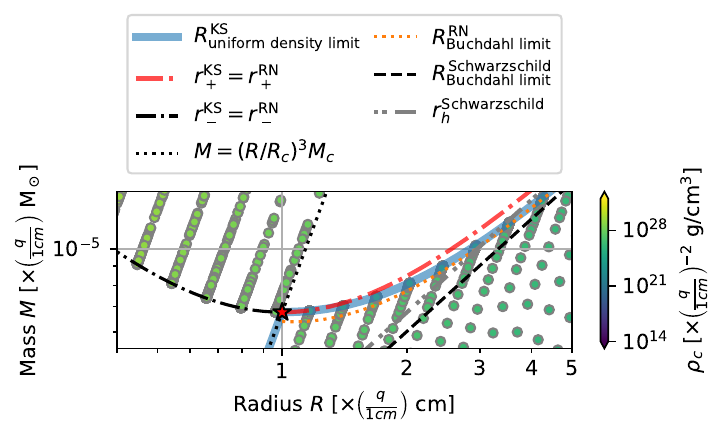}
  }
  \subfigure[\ arbitrary \acsp{eos}]{
    \label{fig:random}
    \includegraphics[width=\columnwidth]{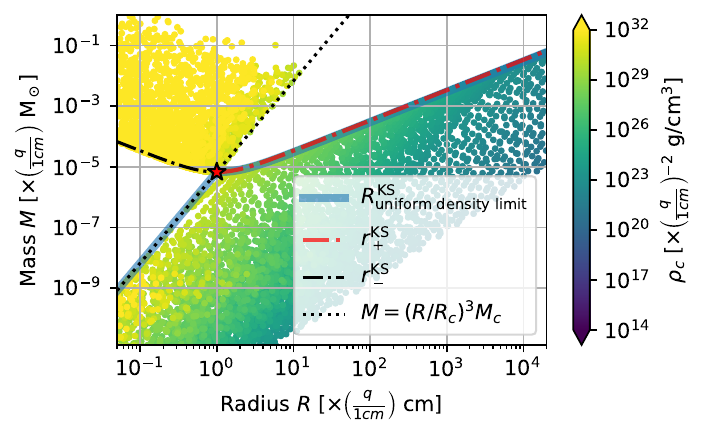}
  }
  \subfigure[\ arbitrary \acsp{eos} (near the minimal black hole)]{
    \label{fig:random:zoom}
    \includegraphics[width=\columnwidth]{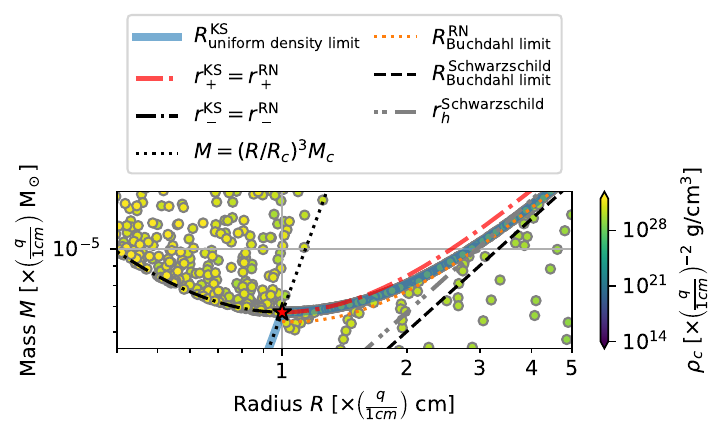}
  }
  \caption{Scatter plots of simulated compact objects \subref{fig:uniform} for uniform density and \subref{fig:random} for arbitrarily monotonic \acsp{eos} with positive pressure are presented. The panels \subref{fig:uniform:zoom} and \subref{fig:random:zoom} are the detailed views near the minimal black hole of the panels \subref{fig:uniform} and \subref{fig:random}, respectively. We see that the compact objects do not exceed the \ac{udl} in both cases, which means that the \ac{udl} is indeed the universal limit of compact objects made of an arbitrary matter, provided that the \acs{eos} is monotonic and the pressure inside the object is everywhere positive. This is consistent with the Buchdahl limit in GR, which is actually the uniform density limit. The objects inside the horizon are new solutions in \ac{hl} gravity, which have not been seen in \ac{gr}. Indeed, in \ac{gr} limit $q \to 0$, the line $M = (R/R_c)^3 M_c$ becomes the vertical axis and the only solutions below the \ac{udl} curve remain. The new solutions are well confined inside the horizon so that they are \textit{de facto} black holes to an observer outside the horizon.
  }
  \label{fig:arbitrary}
\end{figure*}

Furthermore, motivated by the fact that the correction of \ac{hl} to the mass and radius of a neutron star is apparent but that of a white dwarf is negligible even for quite large $q$~\cite{Kim:2018dbs}, we consider a larger $q$ and/or a smaller $R$ such that $R / R_c \sim O(1)$. For this region, the inequality~\eqref{ineq:uniform} reduces to $M < M_c + (1/3q) (R - R_c)^2 - (8/27q^2) (R - R_c)^3+ O \left( q^{-3} (R - R_c)^4 \right)$, which shows that the upper bound of $M$ approaches $M_c$ when $R$ goes to $R_c$. This behavior disappears in \ac{gr} limit, $q \to 0$.

It is interesting to note that the condition $\rho_0 + \beta p > 0$ reduces to the strong energy condition $\rho_0 + 3 p > 0$ in the \ac{gr} limit. However, it is hard to say whether $\beta$ represents the modification of the strong energy condition by \ac{hl} gravity or not.
Instead, we can check the energy condition by introducing the effective stress-energy as $8 \pi \mathcal{T}^\mathrm{eff}_{\mu\nu} \equiv \mathcal{R}_{\mu\nu} - (1/2) \mathcal{G}_{\mu\nu} \mathcal{R}$, then we have
\begin{align}
\rho_\mathrm{eff} &\equiv \frac{1}{8 \pi r^2} \left( r (1-f) \right)' = \frac{\rho_0}{1+\alpha},
  \label{rho_eff} \\
p_\mathrm{eff} &\equiv \frac{1}{8 \pi r^4} \big[ -r^2 (1-f) + 2 r^3 f \Phi' \big] \notag \\
  &= \frac{\rho_0}{3(1+\alpha)} \left( \frac{2 \chi(r)}{\beta \chi(R) - \chi(r)} - 1 \right).
  \label{p_eff}
\end{align}
Now, we see that the effective density and pressure satisfy the strong energy condition,
\begin{equation}
  \label{sec_eff}
  \rho_\mathrm{eff} + 3 p_\mathrm{eff} = \frac{\rho_0}{1+\alpha} \left( \frac{2 \chi(r)}{\beta \chi(R) - \chi(r)} \right) > 0,
\end{equation}
when the \ac{udl} inequality~\eqref{ineq:uniform} holds. Note that the condition
\begin{equation}
  \rho_0 + \beta p = \rho_0 \left( \frac{(\beta -1) \chi (r)}{\beta \chi (R) - \chi (r)} \right) > 0
\end{equation}
has similar form to Eq.~\eqref{sec_eff}, but it is slightly different. This issue may deserve further investigation.

Next, for the critical case of $\alpha = 1$, the curve $M = (4 \pi / 3) \rho_0 R^3 = R^3 / q^2$ is the boundary between the case i) and iii) and is actually the upper bound \eqref{bound:lowmass} of the case i) for $R < R_c$. At this critical density, the solution to the \ac{tov} equation \eqref{eq:p} is given by
\begin{equation}
  \frac{p}{\rho_0} = \frac{\chi(r)}{\chi(R) - \chi(r)},
\end{equation}
from which we can see that the size of a compact object such that $p(R) = 0$ is exactly the same with the radius of the minimal black hole $R = R_c$ and that the pressure is negative for $r < R$. That is, there is no compact object with uniform density in this critical case.

Finally, for the case of $\alpha > 1$, the condition $\rho_0 + \beta p > 0$ holds near the surface of a compact object, since $p(R) = 0$ and $\beta < -3$, while the condition should be negative near the center, provided that the pressure is positive near the center.\footnote{The condition may remain positive $\rho_0 + \beta p > 0$ near center when $\chi(R) > \chi(0)$. However, in this case, the pressure increases with respect to $r$ while $\chi(r)$ decrease to zero, and the derivative of the pressure should suffer discontinuity at the radius where $\chi(r) = 0$ to meet the boundary condition $p(R) = 0$. Thus, this case is excluded in this \paper.} The equality $\rho_0 + \beta p = 0$ holds on the (inner) horizon curve, where $\chi(r) = 0$, which implies that the radius of a compact object with uniform density is larger than the inner horizon and smaller than the outer horizon of a black hole with the same mass, $r_-(M) < R < r_+(M)$. In other words, a compact object with uniform density larger than the average density of the minimal black hole, $\rho_0 > \bar{\rho}_\mathrm{MBH}$, is heavier than the minimal black hole and censored by the outer horizon, which means that the object is a \textit{de facto} black hole to an observer outside the horizon.

\begin{figure*}[tbp]
  \centering
  \subfigure[\ The \acl{ssl} in \acs{hl} gravity]{
    \includegraphics[width=\columnwidth]{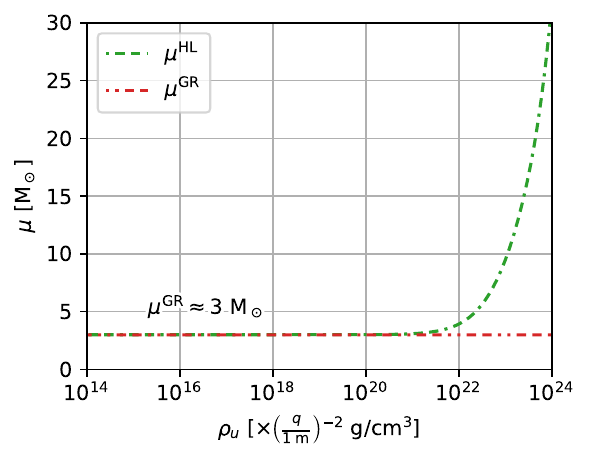}
    \label{fig:causal:factor}
  }
  \subfigure[\ Mass-radius curve of the \acs{ssl} curve]{
    \includegraphics[width=\columnwidth]{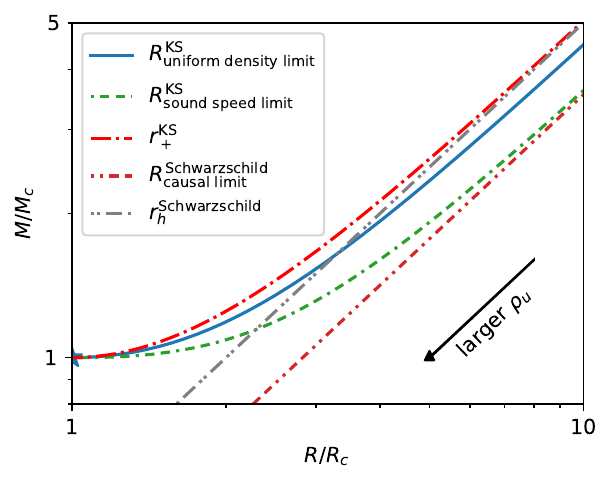}
    \label{fig:causal:mr}
  }
  \caption{\subref{fig:causal:factor} The numerical factor in the causal limit in \acs{gr} becomes a function of the fiducial density $\rho_u$ in the \acs{ssl} in \acs{hl}. 
  \subref{fig:causal:mr} The \acs{ssl} in \acs{hl} is compared to the causal limit in \acs{gr}, considering non-rotating, asymptotically flat vacuum solutions. The black hole horizons and the \acs{udl} of \acs{ks} vacuum also depicted.  All three curves, the horizon, the \ac{udl} and the \ac{ssl}, converge to the minimum of the horizon (star), which represents the minimal black hole.
  }
  \label{fig:causal}
\end{figure*}

\subsection{Arbitrary equation-of-state}

We now consider an arbitrary matter described by a random \ac{eos} and positive pressure. the \ac{tov} equations in HL (\ref{eq:m}) and (\ref{eq:p}) are highly complex, making it difficult to analytically verify if the \ac{udl} is the universal limit for arbitrary matters.

To see if a compact object made of an arbitrary matter is able to exceed the \ac{udl}, we solve the \ac{tov} equations numerically by using the Runge-Kutta method of order 4~\cite{DORMAND198019,Shampine:1986:SPR} implemented in \texttt{SciPy}~\cite{2020SciPy-NMeth}. For simplicity, we introduce dimensionless variables, $\hat{r} = r/q$, $\hat{m} = m/q$, $\hat{\rho} = q^2 \rho$, and $\hat{p} = q^2 p$, then $q$ disappears from the TOV equations. The initial values (central density and pressure) are chosen in the range of $10^{-61} \lesssim \hat{\rho}_c \lesssim 10^{39}$ and $10^{-10} \hat{\rho}_c \le \hat{p}_c \le 10^{10} \hat{\rho}_c$ to cover wide range of $q$. The boundary condition at the surface of the object $\hat{R}$ is given by $\hat{p}(\hat{R}) = 0$, requiring that $\hat{\rho}(\hat{r}) \ge 0$ and $\hat{p}(\hat{r}) \ge 0$ everywhere, $\hat{r} \le \hat{R}$. Furthermore, the density $\hat{\rho}$ is assumed to be decreasing with respect to $\hat{r}$, following Buchdahl.

The simulated compact objects, which are solutions to the \ac{tov} equations, are plotted in Fig.~\ref{fig:arbitrary}. The objects with uniform density are shown in Figs.~\ref{fig:uniform} and \subref{fig:uniform:zoom}: Each dot (circle) represents a compact object characterized by different parameters. It is observed that the objects with uniform density $\rho_0  < 3 / 4 \pi q^2$ are positioned under the \ac{udl} curve (Fig.~\ref{fig:uniform:zoom}), as calculated analytically in the previous section. In contrast, all objects with uniform density $\rho_0 > 3 / 4 \pi q^2$ are located inside the \ac{ks} horizon and form black holes, which is also consistent with the analysis in the previous section.

Next, the simulated compact objects with random \acp{eos} are also shown in Figs.~\ref{fig:random} and \subref{fig:random:zoom} to be formed either under the \ac{udl} curve or inside the \ac{ks} horizon. Each object satisfies its own arbitrary \ac{eos} that is different from each other, which can be seen by the fact that the objects with arbitrary \acp{eos} are not aligned in the order of their central density, while the objects with uniform density are aligned and show a clear gradation in Fig.~\ref{fig:uniform}. Thus, it is shown that the mass of a compact object with an arbitrary \ac{eos} is indeed bounded from above by the \ac{udl}, which turns out to be a universal limit. 

Note that there are many objects with higher central density than the average density of the minimal black hole, $\rho_c \gtrsim 3.2 \times 10^{27} (q/1\ \textrm{cm})^{-2} \ \textrm{g/cm}^3$, under the \ac{udl} curve in Fig.~\ref{fig:random}, compared with Fig.~\ref{fig:uniform}, which shows that a stable compact object can be formed within the \ac{udl}, provided that its average density is less than that of the minimal black hole, $\bar{\rho} < \bar{\rho}_\mathrm{MBH}$, even though its central density is several orders of magnitude higher than $\bar{\rho}_\mathrm{MBH}$.

\section{Sound speed limit}
\label{sec:causal}

In \ac{gr}, the speed of sound $c_s = dp/d\rho$ inside a compact object has to be subluminal, which raises the causal limit given by
\begin{equation}
  M_\text{max} = \mu \left( \frac{\rho_u}{5\times10^{14}\text{ g/cm}^{3}} \right)^{-1/2},
\end{equation}
where $\mu^\text{GR} \approx 3\ \msun$ and $\rho_u$ is the fiducial density that is a typical maximum density described by well known \acp{eos}~\cite{Rhoades:1974fn,Kalogera:1996ci,Astashenok:2021}.

\begin{figure*}
    \centering
    \subfigure[\ $q=4\,000~\textrm{m}$]{
    \includegraphics[width=\columnwidth]{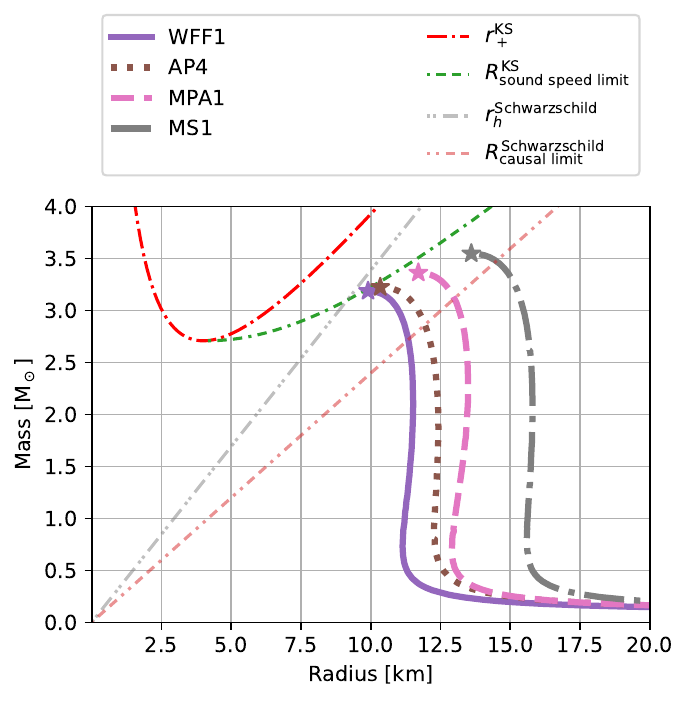}
    \label{fig:mr:q4k}
    }
    \subfigure[\ $q=400~\textrm{m}$]{
    \includegraphics[width=\columnwidth]{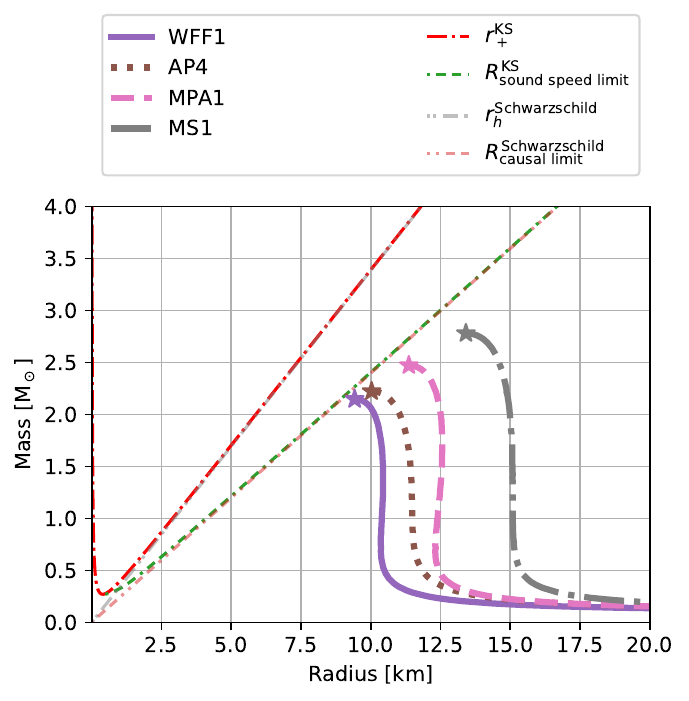}
    \label{fig:mr:q400}
    }
    \caption{Mass-radius relations of the \acsp{eos} selected in the previous work~\cite{Kim:2018dbs} are shown to be below the \acs{ssl} curve for \subref{fig:mr:q4k} $q=4\text{ km}$ and \subref{fig:mr:q400} $q=400\text{ m}$.}
    \label{fig:mr}
\end{figure*}

\begin{table}[tbp]
\begin{tabular}{c | c | c}
  \hline
  \hline
  & $R \sim q$ & $R \gg q$ \\
  \hline
  \acs{udl} &
  $\begin{aligned} 1 &- (R - q) / q \\ &+ (4/3) (R - q)^2 / q^2 \\ &+ O((R - q) / q)^3 \end{aligned}$
  & $\begin{aligned} 4 / 9 &+ (16 / 27) (q / R)^2 \\ &+ O(q / R)^4 \end{aligned}$
  \\
  \hline
  \acs{ssl} &
  $\begin{aligned} 1 &- (R - q) / q \\ &+ (1.014) (R - q)^2 / q^2 \\ &+ O((R - q) / q)^3 \end{aligned}$
  & $\begin{aligned} 0.&354 \\ &+ (6.429\times10^{-6}) (q / R) \\ &+ (0.659) (q / R)^2 \\
                                          &+ (0.2602) (q / R)^3 \\ &+ O(q / R)^4 \end{aligned}$
 \\
  \hline
  \hline
\end{tabular}
\caption{Series expansion of the upper bound of $M / R$ near the minimal black hole ($R \sim q$) and far from it ($R \gg q$) for \acs{udl} and  \acs{ssl}. For \acs{ssl}, the coefficients are fitted by the method of least squares.}
\label{tab:ssl_limit}
\end{table}

Following ref.~\cite{Rhoades:1974fn}, we consider a luminal \ac{eos} for the density larger than the fiducial density to maximize the mass of compact objects. That is, to calculate the \ac{ssl} in \ac{hl}, we solve the \ac{tov} equations with an \ac{eos},
\begin{equation}
  \label{eq:eos:causal}
  p = p_u + (\rho - \rho_u),
\end{equation}
by using the Runge-Kutta method of order 8~\cite{Hairer1993} implemented in \texttt{SciPy}~\cite{2020SciPy-NMeth}, where $p_u$ represents the pressure of a known \ac{eos} at the fiducial energy. Note that it is natural to specify the neutron star \ac{eos} below the fiducial density; however, it contributes negligibly to the crust (or near-crust) of the heaviest compact object that is significant to find the \ac{ssl}. Thus, the \ac{eos}~(\ref{eq:eos:causal}) is used even below the fiducial density in this \paper. The pressure $p_u$ at the fiducial density is set $\rho_u / 100 \le p_u \le \rho_u / 3$, since the pressure of many \acp{eos} --- for example, \acp{eos} in ref.~\cite{Ozel:2016oaf} --- is between $\rho_u / 50$ and $\rho_u / 8$ at $\rho_u=5\times10^{14}\text{ g/cm}^3$. We found that the value of $p_u$ in the above range is negligible in the sense that it only contributes to the mass of the heaviest compact object by $\sim 10^{-7}$. In the limiting cases of near minimal black hole ($R \sim q$) and far from it ($R \gg q$), fitting curves are obtained by using the method of least squares implemented in \texttt{SciPy}~\cite{2020SciPy-NMeth}, and the results are listed in Table~\ref{tab:ssl_limit}, along with the limiting cases of \ac{udl}. Note that the \ac{ssl} is smaller than the \ac{udl} in both the limiting cases, as expected, since \ac{ssl} requires \ac{eos} subluminal, while \ac{udl} is the upper limit of the masses of compact objects with arbitrary \acp{eos}.

The \ac{ssl} in \ac{hl} is depicted in Fig.~\ref{fig:causal}. One might notice that $\mu^\text{HL} \sim \mu^\text{GR}$ for $\rho_u \lesssim 10^{21} (q/1\text{ m})^{-2} \text{ g/cm}^3$ but $\mu^\text{HL}$ grows exponentially for $\rho_u \gtrsim 10^{21} (q/1\text{ m})^{-2} \text{ g/cm}^3$. In \ac{gr} limit, $q\to0$, $\mu^\text{HL}$ becomes $\mu^\text{GR}$ for $\rho_u < \infty$. At $q \sim 1\ \text{m}$ scale, it looks like that the deviation is negligible for $\rho_u \lesssim 10^{21} \text{ g/cm}^3$; however, at $q \sim 5\ \text{km}$, the deviation becomes significant even for $\rho_u \gtrsim 10^{14} \text{ g/cm}^3$ and $\mu^\text{HL}\gtrsim5\ \msun$ for $\rho_u\gtrsim10^{15}\text{ g/cm}^3$ so that it thoroughly covers the lower mass gap between observed black holes and neutron stars. Note that $q \sim 5\text{ km}$ is consistent with the existence of the horizon of a  black hole candidate of mass $\sim 4\ \msun$, GRO J0422+32~\cite{Gelino:2003pr}. Though there exists another estimation of its mass that is $\sim 2.1\ \msun$, which reduces the upper bound of $q$ to $\sim 3 \text{ km}$, there is a chance that it is not a black hole but a compact object~\cite{Kreidberg:2012ud}. Furthermore, in \ac{hl}, an object with a compactness comparable to that of a black hole can form~\cite{Son:2026}.
In this sense, it cannot constrain the upper bound of $q$.

To see if the neutron star is indeed formed within the \ac{ssl}, the mass-radius relations of the selected \acp{eos} for analyses in the previous work~\cite{Kim:2018dbs} are depicted in Fig.~\ref{fig:mr}. The neutron star of maximum mass for each \ac{eos} is marked as a star and is well below the \ac{ssl} curve. All 4 heaviest neutron stars in the case of $q=4\text{ km}$ are heavier than $3\ \msun$, the upper bound in \ac{gr}, which supports that the deviation of $\mu^\text{HL}$ is non-negligible at the scale of $\rho_u\gtrsim10^{14}$. In contrast, those in the case of $q=400\text{ m}$ are below the causal limit in \ac{gr}, since the \ac{ssl} is almost the same at the scale of $\rho_u\lesssim10^{16}\text{ g/cm}^3$.

\section{Discussion}
\label{sec:discussion}

In \ac{hl}, both the \ac{udl} and the \ac{ssl} tend to the horizon near the minimal black hole, where all three curves converge together. This behavior of the \ac{udl} and the \ac{ssl} near the minimal black hole in \ac{hl} explains why the compact objects in \ac{hl} becomes heavier than those in \ac{gr}. Indeed, the masses of the neutron stars with the selected \acp{eos} and the fermionic compact objects that are composed of a fermion with a given mass become larger, while their radii become larger for the neutron stars and smaller for the fermionic compact objects~\cite{Kim:2018dbs,Son:2026}.

Moreover, the fact that \ac{udl} and \ac{ssl} converges at $R=M=q$, where the minimal black hole resides, yields the compactness of a compact object is allowed to be $M / R < 1$, similar to the scalar-tensor gravity, which allows $M / R > 1 / 2$ for $\beta \lesssim 0.2$~\cite{10.1143/PTP.100.291}. Considering a fermion \ac{eos}, it has been shown that the compactnesses of fermionic compact objects becomes larger than those in \ac{gr} and the compactness gap between black holes and other compact objects, witnessed in \ac{gr}, vanishes in \ac{hl}~\cite{Son:2026}. For example, for $q \sim 4\ \mathrm{km}$, a compact object of a fermion of mass $\sim 1\ \mathrm{GeV}$ can have the compactness $M / R \gtrsim 1 / 2$ and fermions of mass $\sim 2\ \mathrm{GeV}$ can form a compact object of $M \sim q \sim R$. (for details, see ref.~\cite{Son:2026}.) For much larger objects with $R/q \gg 1$, of course, the \ac{udl} and \ac{ssl} curves converge to the Buchdahl and the causal limit curves of Schwarzschild vacuum, respectively.

The \ac{ssl} obtained in this \paper{} does not represent the definitive theoretical limit due to the violation of Lorentz symmetry in \ac{hl} gravity so that the speed of light is not a universal constant $c$. To resolve this, the speed of light in \ac{hl} gravity and its relation to the speed of sound inside a compact object should be studied. However, it is beyond our scope that is to show the \ac{ssl} tends to the horizon and eventually converges with the horizon at its minimum. Furthermore, the heavier compact objects should be allowed when the limit on the speed of sound is relaxed, which yields that the actual \ac{ssl} curve should be closer to the horizon curve and the deviation might be seen in even lower fiducial density scale. This issue deserves further investigation.


\section*{Acknowledgements}
I would like to thank M.-I.\ Park, G.\ Kang and J.\ Kim for the helpful discussion.
This work was supported by the National Research Foundation of Korea (NRF) grant funded by the Korea government (MSIT) (No.\ 2021R1A2C1093059).


\end{document}